\documentclass[         %
aps,                    
jpa,                    
showpacs,               
nofootinbib,            
showkeys,               %
preprintnumbers,        %
floatfix]               
{revtex4}               
\tighten
\draft
\usepackage{graphicx} 

\begin{document}

\title{Neutrino interactions in astrophysics and the third neutrino mixing angle $\theta_{13}$}

\pacs{13.15.+g, 14.60.Lm, 25.30.Pt, 26.30.Jk, 95.30.Cq}
\keywords      {Neutrino interactions, neutrino mixing, neutrino astrophysics}

\author{A.B. Balantekin}
\address{Physics Department, University of Wisconsin, Madison, WI  53706 USA}

\begin{abstract}
The third neutrino mixing angle $\theta_{13}$  is the least known one. In this contribution, after a brief discussion of the current efforts to determine the value of this angle better,  various astrophysical implications of a non-zero value of $\theta_{13}$ are summarized.

\end{abstract}

\maketitle


\section{Introduction}
We learned a lot about neutrino-mixing from recent experiments, yet our current knowledge 
of the third mixing angle, $\theta_{13}$, is very limited. The CP-violating phase $\delta$ 
in the neutrino mixing matrix  
\begin{equation}
 {\bf T}_{23}{\bf T}_{13}{\bf T}_{12}  = 
\left(
\begin{array}{ccc}
 1 & 0  & 0  \\
  0 & C_{23}   & S_{23}  \\
 0 & -S_{23}  & C_{23}  
\end{array}
\right)
\left(
\begin{array}{ccc}
 C_{13} & 0  & S_{13} e^{-i\delta}  \\
 0 & 1  & 0  \\
 - S_{13} e^{i \delta} & 0  & C_{13}  
\end{array}
\right) 
\left(
\begin{array}{ccc}
 C_{12} & S_{12}  & 0  \\
 - S_{12} & C_{12}  & 0  \\
0  & 0  & 1  
\end{array}
\right) ,
\end{equation}
where $C_{ij} = \cos \theta_{ij}$, $S_{ij} = \sin \theta_{ij}$, is likely to be observable only for relatively large values of $\theta_{13}$.  Hence it is crucial to have a good idea about the value of this angle if we want to experimentally look for the CP-violation in the neutrino sector, and theoretically examine consequences of breaking this symmetry, such as exploring the origin of  the matter-antimatter symmetry in the Universe.  In this talk, various implications of a non-zero value of 
$\theta_{13}$ will be briefly summarized.

\section{Current limits on $\theta_{13}$}

Current experimental limit  on the mixing angle $\theta_{13}$ suggested by the Particle Data Group is $\sin^2 (2\theta_{13}) < 0.19$ \cite{Amsler:2008zzb} at 90\% C.L., based primarily on the Chooz experiment. However, three reactor neutrino experiments, Double Chooz \cite{Ardellier:2006mn}, Daya Bay \cite{Guo:2007ug}, and RENO \cite{:2010vy}, which currently are at different stages of construction, aim to measure this mixing angle. A measurement of $\theta_{13}$ by observing the electron neutrino appearance is also one of the main goals of the T2K experiment \cite{Ichikawa:2010zz}.  The current T2K run partially coincident with this conference is expected to reach to a sensitivity exceeding the current Chooz limit. 

After the publication of the most recent KamLAND results  \cite{:2008ee} it was noted that 
best fit values for the neutrino mass difference and mixing angle determined from the KamLAND and solar neutrino data in the two-flavor analysis are not the same \cite{Balantekin:2008zm}. Even though previously undiscovered interactions that distinguish neutrino (those coming from the Sun) and antineutrino (those coming from a reactor) masses and mixings could cause such a difference, the  
simpler explanation is a non-zero value of $\theta_{13}$. As shown in the Figure 1, finite values of 
$\theta_{13}$ should place the best fit values of the of solar and reactor neutrino data at different locations. Analyses by many authors indeed point out to a non-zero value of $\theta_{13}$ 
(see references \cite{Fogli:2008ig} through \cite{GonzalezGarcia:2010er}). 
Care must be taken, however as errors given are non-Gaussian when a non-zero value of $\theta_{13}$ is quoted.  Most recent low-threshold analysis of both phases of the 
Sudbury Neutrino Observatory (SNO) suggests $\sin^2 \theta_{13} < 0.057$ at 95\% C.L. with a best fit value 
of $2.00^{+2.09}_{-1.63} \times 10^{-2}$ \cite{Aharmim:2009gd}. One should note that it is not difficult to introduce theoretical  models with such large values of $\theta_{13}$ \cite{Albright:2008rp,Goswami:2009yy}. 
 
\begin{figure}
\includegraphics{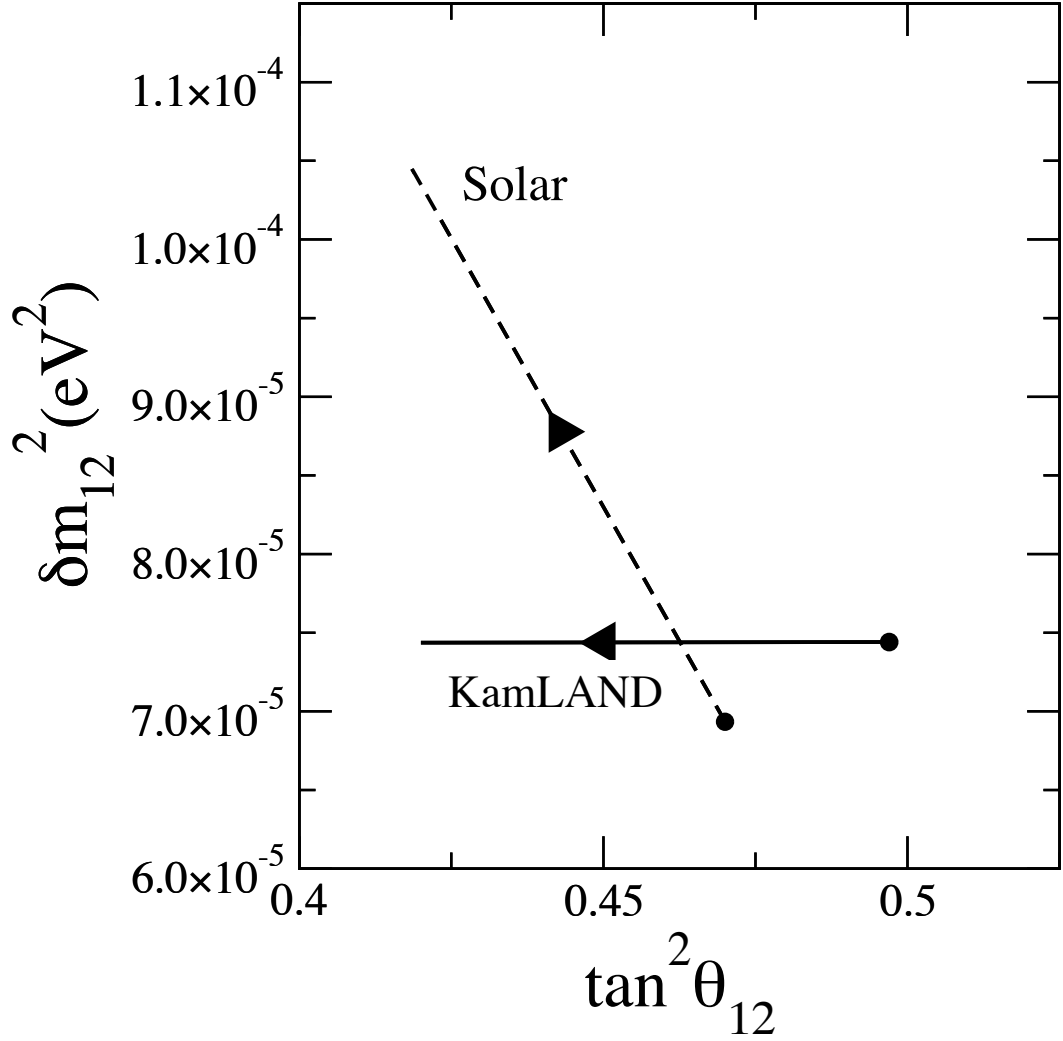}
 \caption{The change in the best fit values of $\delta m_{12}^2$ and $\theta_{12}$  with 
increasing value of $\theta_{13}$. Parameter values corresponding to 
$\theta_{13} = 0$ are indicated by filled circles. Both for KamLAND and 
solar neutrino experiments the range $0 \le \sin^2 2 \theta_{13} \le 0.1$ 
is shown \cite{Balantekin:2008zm}.}
\end{figure} 
 
One note of caution is in order, namely that the value of $\theta_{13}$ is small. Theoretical input into the analysis of neutrino data includes other small quantities which are not well-determined either. 
For example 
neutrino-deuteron cross sections are a crucial input into the analysis of the SNO data. One approach to the calculation of neutrino-deuteron cross sections from the first principles would be using the effective field theory: In this reaction, which is below the pion threshold,  the $^3S_1\rightarrow ^1S_0$ transition dominates and one only needs to determine the coefficient of the two-body counter term, ${\rm L}_{\rm 1A}$, associated with the isovector two-body axial current, to predict the cross-section  
\cite{Butler:1999sv,Ando:2002pv}. 
The  ${\rm L}_{\rm 1A}$  term can be obtained by either comparing the cross section $\sigma(E) = \sigma_0(E) + {\rm  L}_{\rm 1A} \sigma_1(E)$ with the cross-sections calculated using other approaches 
(save the model dependences \cite{Mosconi:2007tz}) or measured experimentally 
(e.g. using solar neutrinos as a source). It can be shown that uncertainties coming from the lack of knowledge of $\theta_{13}$ and the counter-term ${\rm L}_{\rm 1A}$ in fitting the solar neutrino flux data are comparable \cite{Balantekin:2004zj}.
In addition, signatures for non-zero values of $\theta_{13}$ may also mimic signatures of physics beyond the Standard Model \cite{Palazzo:2009rb}. 

A good recent summary the current status and near-term prospects of $\theta_{13}$ phenomenology is given in Ref. \cite{Mezzetto:2010zi}.

\section{$\theta_{13}$ and core-collapse supernova physics}

Understanding the dynamics of core-collapse supernovae is important not only to better understand later stages of the stellar evolution, but also to ascertain the suitability of core-collapse supernovae as possible sites of the r-process nucleosynthesis. For nuclei with A>100, r-process abundances observed in the solar system and iron-poor stars agree very well, suggesting a universal site. 

Of course understanding the supernova dynamics is not the whole story. 
To understand the r-process one also needs to understand beta-decays of nuclei both at and far-from stability:  Half-lifes at the r-process ladder nuclei (N = 50, 82, 126), where abundances peak, should be 
known well. One also needs accurate values of the energies of the initial and final states as well as the matrix elements of the Gamow-Teller operator ${\bf \sigma} {\bf \tau}$, (even the forbidden operators  
${\bf r \sigma} {\bf \tau}$) between those states.  Understanding the spin-isospin response of a broad range of nuclei to a variety of probes is indeed crucial for a wide range of astrophysics applications!  

Yields of r-process nucleosynthesis are determined by the electron fraction, or equivalently by the neutron-to-proton ratio, $n/p$. Interactions of the neutrinos and antineutrinos streaming out of the core both with nucleons and seed nuclei determine the n/p ratio. Hence to understand all the physics 
associated with the core-collapse supernovae, it is crucial to understand neutrino properties and interactions \cite{Balantekin:2003ip}. Before those neutrinos emitted from the core reach the r-process region they undergo matter-enhanced neutrino oscillations as well as coherently scatter over other neutrinos.  Many-body behavior of this neutrino gas is still  being explored, but is likely to have significant impact on r-process nucleosynthesis. Possible existence of exotics, such as sterile neutrinos,  would also impact supernova 
dynamics \cite{Fetter:2002xx}. 

The MSW potential is provided by the coherent forward scattering of electron neutrinos off the  electrons in dense matter via a W-exchange.  (There is a similar term with Z-exchange. But since it is the same for all neutrino flavors, it does not contribute at the tree level to phase differences unless we invoke a sterile neutrino). 
If the neutrino density itself is also very high then one has to consider the effects of neutrinos scattering off other neutrinos. This is the case for a core-collapse supernova. Both the O-Ne-Mg-core collapse and iron-core collapse scenarios should be effected by neutrino-neutrino interactions. 

For the illustrative case of two flavors of neutrinos (and no antineutrinos), forward scattering of neutrinos from other neutrinos is described by the Hamiltonian \cite{Balantekin:2006tg} 
\begin{equation}\label{2}
H_{\nu \nu} = \frac{\sqrt{2} G_F}{V} \int d^3p \> d^3q \>  (1-\cos\vartheta_{pq}) 
\> {\bf J}(p) \cdot {\bf J}(q) .
\end{equation}
where $\cos\vartheta_{pq}$ is the angle between momenta $p$ and $q$. In Eq. (\ref{2}) the "neutrino spin" operators are 
\begin{equation}\label{1}
J_+(p)= a_x^\dagger(p) a_e(p), \ \ \ \
J_-(p)=a_e^\dagger(p) a_x(p), \ \ \ \
J_0(p)=\frac{1}{2}\left(a_x^\dagger(p)a_x(p)-a_e^\dagger(p)a_e(p)
\right) . 
\end{equation}
In Eq. (\ref{1}), "neutrino spin" operators are written in terms of the flavor creation and annihilation operators. 
It is possible to include all three flavors and antineutrinos in the equations above. Many authors studied 
neutrino propagation with neutrino-neutrino interactions (see references \cite{Qian:1994wh}
through \cite{Duan:2008fd}).
One typically finds large-scale, collective flavor oscillations deep in the supernova envelope, which is sensitive to the value of $\theta_{13}$.  Of particular note is the presence of spectral swaps between 
electron neutrinos and neutrinos of other flavors, which seem to persist even for very small (but non-zero) values of $\theta_{13}$ (see references \cite{Duan:2008za} through \cite{Dasgupta:2010cd}). 
Especially for iron-core supernovae, those swap features are present 
at somewhat lower (i.e. solar neutrino) energies than one usually associates with supernovae. 

Since the neutron-to-proton ratio, $n/p$, or alternatively the electron fraction, is determined by the  interactions and the environment-dependent propagation of the neutrinos and antineutrinos streaming out, the value of $\theta_{13}$  in an important quantity in determining abundances of r-process elements.   Preliminary work suggests that for larger values of $\theta_{13}$, neutrino-neutrino interactions may hinder the r-process by increasing the electron fraction \cite{Balantekin:2004ug} although much work remains to be done.

The value of $\theta_{13}$ may also effect light-element production through the so-called  $\nu$-process. Yoshida {\it et al.} found that the the cosmological abundance ratio $N(^7Li)/N(^{11}B)$ not only depends on $\theta_{13}$, but this dependence is different for the normal and inverted mass hierarchies \cite{Yoshida:2006sk}.





\section*{Acknowledgments}
This work was supported in part 
by the U.S. National Science Foundation Grant No. PHY-0855082 
and 
in part by the University of Wisconsin Research Committee with funds 
granted by the Wisconsin Alumni Research Foundation.

\end{document}